# Jamming Detection for IR-UWB Ranging Technology in Autonomous UAV Swarms

Pavlo Mykytyn[1,2], Marcin Brzozowski[1], Zoya Dyka[1] and Peter Langendoerfer[1,2]
[1] IHP - Leibniz-Institut für innovative Mikroelektronik, Frankfurt (Oder), Germany
[2] BTU Cottbus-Senftenberg Cottbus, Germany
{brzozowski, dyka}@ihp-microelectronics.com, {pavlo.mykytyn, peter.langendoerfer}@b-tu.de

*Abstract*— **Jamming is a form of the Denial of Service (J-DoS) attack. It is a significant threat that causes malfunction in Unmanned Aerial Vehicle systems, especially when used in hostile environments. The attackers mainly operate in the wireless communication environment by following a few preexisting scenarios. In this paper, we propose an idea for a Jamming detection mechanism. The mechanism utilizes the network parameters available to the system and some additional measures to distinguish between bad transmission quality and Jamming to avoid false positive alarms. After detecting a Jamming attack, appropriate countermeasures or mitigation techniques can be applied to keep the system safe.**

## I. INTRODUCTION

Digitalisation is happening very quickly and enables new innovative solutions. Unmanned Aerial Vehicles (UAVs) are used more and more frequently in industrial logistics and delivery [1], monitoring of large areas [2] and in agricultural field management [3]. UAVs face challenges when they shall be used in complex and time expensive tasks. For example, in agriculture UAV monitoring helps to early detect and combat various diseases of plants, as well as fungal and insect infestations. Conventional methods include spraying the entire field with pesticides or chemicals, which impacts the quality and sustainability of the end product majorly. That could be prevented by using UAVs for a local decision making and treatment of plants. This can save chemicals and improve product quality significantly.

According to the current state of the art and technology, UAV swarms face the following challenges :

1. Time of flight limited to 20-25 min.
2. Covering large areas
3. Collision avoidance
4. Cyber-attacks

To tackle the first challenge, a landing platform at the base station can be placed for a quick and seamless battery change. On the platform itself a UWB (Ultra-Wideband) technology can provide fast and precise distance measurements to land a UAV onto the charging platform. This process interrupts the mission for only a few minutes and allows smooth operation. After replacing the battery, the UAV returns to the position where its mission was paused and proceeds until it is completed. This allows concentrating on the mission, especially if the battery changing platform is portable and can be placed close to the location where the mission is carried out.

To successfully overcome the second challenge, i.e. covering large areas, a swarm formation has to be able to perform complex, sequential or parallel tasks during the flight. According to the current planning, up to 16 UAVs could be involved at the same time for such applications. This would significantly reduce the task execution time.

UWB and Radar technologies can be applied as means for safe navigation and collision avoidance. The IR-UWB ranging technology has the following advantages compared to Narrowband technologies: a higher multipath resolution, higher data transfer rates, low probability of detection and interception, low latency and high accuracy. These advantages make IR-UWB especially suitable for the UAV applications. For avoiding collisions a smart collision avoidance algorithm can be developed based on the GPS positioning, inertial navigation system and inertial measuring unit of the UAV and IR-UWB distance ranging technology.

Recently, many cyber- and physical attacks were reported to happen [4]. Jamming attacks are a kind of intentional (malicious) interference attacks. To prevent them, a Jamming detection mechanism has to be implemented in the UAVs. After a quick and consistent detection of Jamming attacks proper countermeasures have to be applied to prevent any damage of UAVs or surroundings.

The rest of the paper is organized as described below. Section II discusses the Jamming problem if IR-UWB technology is used: we give a brief description of UWB technology and common types of Jammers. Statistics often used for Jamming detection mechanisms are explained in Section III. Section IV gives a short overview of related work. Section V presents our idea of a Jamming detection mechanism for the IR-UWB ranging technology. The paper is concluded in Section VI.

## II. IR-UWB AND JAMMING

Collision avoidance algorithms require the knowledge about the distance between UAVs in a swarm. One of the main concerns during an autonomous UAV swarm flight is a precise, fast and reliable distance measurement between the UAVs . If the distance measurements are distorted or impossible to perform due to a Jamming attack, it could potentially lead to a collision and damaging the UAVs and /or surroundings. Therefore to provide robust, fast and precise distance measurements IR-UWB ranging technology can be used.

### A. Ultra-Wideband ranging technology

Ultra-Wideband signal is formally any signal that has a bandwidth of 500 MHz or more. UWB ranging technology discussed in this work is based on the Impulse Radio architecture. IR-UWB ranging or communication is a transmission of binary data, using low energy short narrow pulses or bursts (in the order of nanoseconds) over a wide spectrum of frequencies with a high time resolution. According to the EU regulations UWB frequency band is defined within 6.0 - 8.5 GHz with a maximum spectral power density of -41.3 dBm/MHz.

The ability to measure the distance between two devices with a precision of a few centimeters is an important advantage for many location-aware applications in indoor and outdoor scenarios. Many of these applications are security-sensitive requiring reliable distance measurements, even in the presence of an adversary interfering with the ranging process [5]. In that matter, UWB ranging technology has the following advantages over the Narrowband applications:

- **High data rate transmission:**
  Due to the wide bandwidth up to a few GHz, UWB systems can support 500 Mb/s or even higher data transmission rates in the range of 10 m.

- **Low energy consumption:**
  Due to the low transmission power, energy consumption remains low, which is good for energy constraint applications.

- **High precision ranging:**
  Because of the short duration of nanosecond pulses, UWB systems have good time-domain resolution, low latency (<1ms) and can provide centimeter accuracy for location and tracking applications.

- **Multipath robustness:**
  Due to the good penetration properties UWB systems have good multipath performance even in dense multipath environments.

- **Security:**
  Due to the low power spectral density and operation below the noise floor, it is extremely difficult for unintended users to detect UWB signals. That is why the probability of interception is low as well. The interference resistance of UWB technology is also higher than the resistance of Narrowband applications, due to the bandwidth of UWB.

Despite its immunity to detection and interception, dedicated attackers can still degrade the performance of UWB transmission or completely disrupt it using a strong source of Narrowband signal close to the transciever of UWB signals. The main disadvantage of UWB ranging technology is its limited range. The power restriction limits UWB operation to about 20-30 meters at around 20-50 Mbps, but other configurations with shorter range and higher transmission rate or longer range and lower transmission rate are also possible.

### B. Types of Jammers

Jammers in Wireless Sensor Networks (WSNs) are malicious wireless nodes that cause intentional interference. There are various attack strategies that a Jammer can perform in order to interfere with the WSN. As a consequence these attacks will have different levels of effectiveness, and may also require different detection strategies. From a practical point of view, four common Jamming strategies, that have proven to be effective in disrupting wireless communication, are considered in [6]. They are namely: Constant Jammer, Deceptive Jammer, Random Jammer and Reactive Jammer (also classified as elementary Jammers in [7]).

- **Constant Jammer:** Emits random bits of uninterrupted signal into the environment following no protocol rules. The communication between the nodes can get disrupted by these activities as long as the jamming signal is emitted. This type of Jammers is very inefficient since they have to emit a high energy signal at all times.

- **Deceptive Jammer:** Instead of emitting random bits of signal constantly, deceptive Jammers transmit legitimate packets into the environment at high rates. In this case, receiver nodes remain constantly busy and communication is rendered unavailable. Since the attack happens persistently, the deceptive Jammer is also not energy efficient.

- **Random Jammer:** This type of Jammer randomly emits constant bits of a signal for some amount of time and stays in sleep mode for the rest of the time. Random Jammer can imitate constant or deceptive Jammers and is slightly more energy efficient than those.

- **Reactive Jammer:** Always listens to the communication. In case the communication is initiated in the environment, it starts the attack. Legitimate packets sent out by sensor nodes get corrupted by the signal emitted by the Jammer. Because reactive Jammers listen for the communication constantly, they are not very energy efficient either. Although, they are much harder to be detected than the other types.

Fig.2 represents a Jammer ($J_X$) interfering with an ongoing IR-UWB transmission between the transmitter ($T_X$) and the receiver ($R_X$). The graph represents the frequency ($f_X$) and signal power that could be used by the adversary, but exceeding the one regulated by standards.

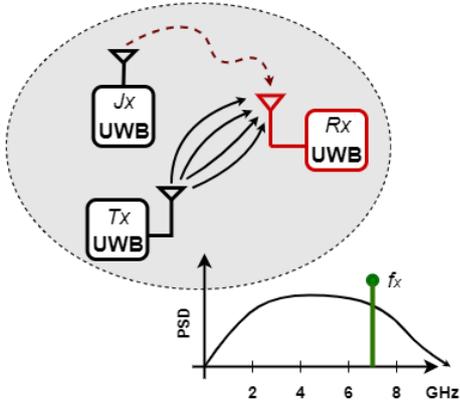

Fig.2: One of the Jamming scenarios.

## III. STATISTICAL PARAMETERS FOR JAMMING DETECTION

Jamming detection mechanisms are very important for a safe UAV operation. However it could be challenging to differentiate a Jamming scenario from the legitimate transmission scenarios, especially under congested network conditions. In WSNs, statistical Jamming detection mechanisms are based on the parameters that represent the transmission quality. Yet parameters like RSSI or SNR for instance, cannot be used as a Jamming detection statistics for the IR-UWB transmission due to the specific characteristics and nature of the UWB signal. Commonly used statistics for Jamming detection mechanisms in WSNs are presented in the rest of this section.

- **Packet Delivery Ratio** (PDR) is the ratio of the number of packets delivered to the total number of packets sent from a source node to a destination node in the network [8]. It is usually measured in percent. The value can fluctuate depending on the network conditions, but even in congested scenarios the PDR for UWB is usually higher than 75%.

$$PDR = \frac{Number\ of\ correctly\ delivered\ packages}{Total\ number\ of\ transmitted\ packages} 100\% \quad (1)$$

- **Bit Error Ratio** (BER) is the number of bit errors divided by the total number of transferred bits during a studied time interval. Bit error ratio is a unitless performance measure, sometimes expressed as a percentage. In use, the number of errors are counted and presented as a ratio such as 1 in 1,000,000.

$$BER = \frac{Number\ of\ bit\ errors}{Total\ number\ of\ transferred\ bits} \quad (2)$$

- **Bad Packet Ratio** (BPR) is the ratio of the number of erroneous packets to the number of total packets received by a node. The PDR and BPR parameters determine the quality of communication for transmitter side and receiver side respectively. These two parameters are inversed in most cases, but in some cases both the BPR and PDR values can be low.

$$BPR = \frac{Number\ of\ erroneous\ packets}{Total\ number\ of\ received\ packets} \quad (3)$$

- **Packet Send Ratio** (PSR) is the ratio of packets that are successfully sent out by a transmitter compared to the number of packets it intended to send out at the MAC layer [6].

$$PSR = \frac{Number\ of\ succesfully\ sent\ packets}{Number\ of\ the\ packets\ intended\ to\ send} \quad (4)$$

## IV. RELATED WORK

Various Jamming detection mechanisms for WSNs presented in the literature were studied in this work. Many jamming detection mechanisms involve recording network transmission parameters in an attack-free environment to define their initial values and set the thresholds for attack-free communication. The measured transmission parameter values are then compared to the thresholds for Jamming detection, for example in [9-13]. These methods fall in the category called statistical anomaly detection. Although a lot of Jamming detection mechanisms for WSNs are published in the literature, only some of them are applicable for the IR-UWB ranging technology. Table 1 displays a short overview of these mechanisms and the statistics they use for Jamming detection.

TABLE I. AN OVERVIEW OF DETECTION MECHANISMS

| Name of mechanism | Type of detection | Detection statistics |
|---|---|---|
| QUJDA basic [9] | Anomaly based | PDR, BPR, ECA |
| QUJDA advanced [9] | Anomaly based | PDR, BPR, ECA |
| Fast jamming detection [10] | Anomaly based | PDR |
| PDR with RSSI consistency checks [11] | Anomaly based | PDR, RSSI |
| PDR with GPS consistency checks [11] | Anomaly based | PDR, GPS |
| PDR with consistency checks and PSR [12] | Anomaly based | PDR, RSSI/GPS, PSR |
| Reactive Jammer detection [13] | Anomaly based | BER, RSSI |

In [9] authors proposed two detection mechanisms – basic and advanced. The first one is called QUJDA (Query - based Jamming detection algorithm) that is an anomaly-based approach. The idea of the approach is to compare the initial network conditions with the current network conditions, using the available network parameters such as PDR, BPR and ECA. The threshold values are defined by the initial tests and then set to detect abnormal conditions. The second mechanism is based on QUJDA, but includes the information from the neighboring nodes. If an abnormal network condition is identified, the first node sends an inquiry on the network conditions to its neighboring node. If the neighbor node confirms abnormalities, a Jamming attack is declared as detected. The biggest drawback of both these methods is a long processing time required due to many parameters.

In [10] authors propose Jamming detection using just the PDR values. The method declares a Jamming attack or raises suspicion of an attack as soon as the PDR value drops significantly. However this approach does not include congested network or weak link scenarios as a part of real world communication between the nodes.

In [11] authors developed two methods based on PDR along with consistency checks for detecting four types of Jammers. Authors conclude that a single measurement cannot alone determine the presence of a Jammer accurately and efficiently. Therefore, the first method is based on measuring PDR and comparing it with the preset threshold value. Low PDR would normally indicate Jamming, but it can also be due to several other factors like network congestions or weak signal. To confirm a low PDR is due to Jamming – RSSI consistency check is introduced along with PDR. If the PDR value is low but RSSI is high it indicates Jamming. If both of them are low it indicates a congested link or weak signal. The downside of using this method for the IR-UWB technology is that in the UWB transmission it is hard to measure RSSI due to its signal properties. In the second mechanism along with PDR author proposed location consistency check. If the node detects low PDR value it then checks on the PDR values with its nearest neighbors using their GPS coordinates. If the neighboring nodes confirm to have low PDR as well, a Jamming attack is then concluded.

In [12] authors improve the method developed in [11] by adding the Packet Sent Ratio (PSR) metric along with PDR at the transmitter side to be able to differentiate types of Jamming attacks.

In [13] authors propose a new method based on BER and RSSI values. This approach focuses on the individual bit errors within a packet and estimates whether the bit errors occurred in the packet were due to a Jamming attack or just a weak/congested communication link. Whenever a node receives a packet, it also records the RSS for each received bit in the packet. After looking at the number of bit errors and at the respective RSS value sampled during the reception of this bit, it estimates if the bit error was due to the external interference (high RSS) or if the error was due to a weak signal (low RSS). The downside of this approach for UWB technology is the problem of precise RSS measuring, due to the low power spectral density of the signal. Therefore, it renders this approach ineffective.

In [14] authors experimented with a multi-user environment of IR-UWB systems. As a results the authors indicate that for proper multi-user experience it is important to select the right repetition rate separation. Otherwise, the closer the repetition rate of the second transmitter gets to the repetition rate of the first transmitter, the worse the dynamic range becomes. This is especially useful for the case when a Jammer uses another UWB device to distort the communication. Authors also tested the effect of the narrowband interference on IR-UWB transmission and noted that narrowband interferers influence the operation of an IR-UWB system only in close vicinity to the UWB transmission, which drastically reduces the probability of a conflict. However Jamming of a transmission is still possible when two interferers are spaced with a difference of the UWB impulse repetition rate, but in all cases of narrowband interference, a simple change of the repetition rate can easily resolve the IR-UWB transmission.

V. PROPOSED DETECTION MECHANISM

IR-UWB ranging is a suitable technology for distance measurements, due to its high robustness and low latency. For Jamming detection a simple, easily executable, fast and effective mechanism is required, as in autonomous UAV swarms real time behavior and computational speed are very crucial. These are the reasons why our Jamming detection mechanism is based on statistics. Unlike the other statistical threshold-based mechanisms, our mechanism does not use a single static predefined threshold value for every case, but rather a variable threshold by applying a pre-defined function.

As the main statistical parameter for Jamming detection mechanism, we selected the Packet Delivery Ratio. However, using a single statistical parameter is not sufficient for an accurate Jamming detection because several other reasons like a congested network or weak signal can also cause low PDR. Therefore, we selected the distance $d$ between the transmitter and the receiver pair as an additional detection parameter. This parameter helps avoiding false Jamming detections if a large distance between the nodes causes a weak transmission link and – consequently – low PDR values.

The distance between the UAVs influences the link quality notably. For instance, if the distance between the UAVs increases, the signal quality will gradually degrade, which would naturally cause lower PDR. Thus, to eliminate false detections, the PDR threshold (denoted further as $PDR_{thr}$) has to be set appropriately. A constant (invariable) PDR threshold value does not appropriately reflect the link quality at different distances. Thus, we decided to adjust the $PDR_{thr}$ depending on the distance between the UAVs. To be able to adjust the PDR threshold, we use a relation between the distance $d$ and the PDR threshold $PDR_{thr}$ based on experimental measurements made beforehand. The measurements are taken starting from the minimal operational distance and finishing at the maximal operational distance recording both PDR and distance values. Analyzing the change in PDR by increasing the distance we adjust $PDR_{thr}$ for each distance measurement accordingly. After the measurements are completed and the dependence between

$PDR_{thr}$ and distance is defined we obtain the rest of the values that were not covered experimentally using interpolation. Using this data set we define the relation between $PDR_{thr}$ and distance in form of a function $PDR_{thr}=f(d)$.

The main idea of our Jamming detection mechanism is to measure the current distance between the UAVs, and the current PDR values simultaneously. The measured PDR value is then compared to its threshold value expected for the current distance. If the distance $d$ between the nodes did not exceed the operational maximum $d_{max}$, and the current PDR value is lower than the expected threshold $PDR_{thr}=f(d)$ for the current distance $d$, a Jamming attack is declared as detected. By practical applications, it is not excluded that the distance $d$ can exceed the pre-defined operational maximum distance, i.e. $d > d_{max}$. Using the pre-defined function $PDR_{thr}=f(d)$ it is always possible to obtain the PDR threshold for each distance, even for large distances. But in this special case it indicates that the nodes are too far apart. The probability that a low PDR is due to the weak signal and not to Jamming is very high. Thus, we denote this special case as an exceptional case and classify it as "no Jamming" even if the currently measured PDR is lower than the threshold calculated for the distance. We used the following relations for PDR and distance measurements for the Jamming detection:

- $d < d_{max}$ and $PDR < PDR_{thr}(d) \rightarrow$ Jamming, because the currently measured PDR is lower than its threshold within the operational distance;

- $d < d_{max}$ and $PDR > PDR_{thr}(d) \rightarrow$ no Jamming, because the currently measured PDR is higher than its threshold within the operational distance, i.e. it is a regular state, no concerns for Jamming;

- $d > d_{max}$ and $PDR < PDR_{thr} \rightarrow$ no Jamming, it is an exceptional case: the currently measured distance is higher than the operational maximum distance. This large distance is a plausible reason for the weak signal and the small PDR value, i.e. the probability of Jamming is low.

The maximal operational distance can be set experimentally and saved. PDR values as well as the distance $d$ between the nodes should be measured regularly. Ideally at least once per second or more frequently.

Algorithm 1 presents the pseudocode of our Jamming detection mechanism. The maximal operational distance $d_{max}$ is pre-defined, as well as the function for obtaining the $PDR_{thr}$. For each currently measured PDR and distance $d$ the PDR threshold will be obtained using the pre-defined function $PDR_{thr}=f(d)$. The next step is comparing the measured PDR value to the obtained threshold value as well as the currently measured distance to the maximal operational distance. If a Jamming attack is detected, appropriate countermeasures are run to keep the UAV safe. Otherwise the regular UAV operation continues until the next measurement.

**Algorithm 1.** Jamming detection mechanism

```
Jamming_Detection()
{
 pre-defined: d_max, PDR_thr=function(d)
 get values(PDR,d);/measuring PDR and d

 if (PDR<PDR_thr AND d<d_max)
    JAMMING=TRUE;   /*low PDR within operational
                      distance*/
 else   /* a regular state: PDR>PDR_thr by d<d_max,
          or an exceptional case: PDR<PDR_thr by d>d_max*/
    JAMMING=FALSE;
 end if
 return JAMMING;
 }
```

## VI. Conclusion and future work

Jamming attacks are a serious threat to safe operation, especially if devices with limited resources are attacked, for example UAVs in a swarm. In this paper, we have studied the issue of detecting the presence of a Jamming attack and examined the ability of different statistics to detect the presence of a Jammer in WSNs. We also proposed a Jamming detection mechanism for an IR-UWB ranging technology in an autonomous UAV swarm. The proposed mechanism can differentiate a Jamming attack scenario caused by various Jammers from a week signal scenario thanks to an adjustable threshold value. Another advantage of this detection mechanism is that it does not require any additional hardware to be implemented and is simple, straightforward and fast. This makes the approach applicable for real time UAV operation. It should also be mentioned that this is just the initial stage of the research and a series of experiments and implementations into an autonomous UAV swarm are planned in the near future, as well as performance testing of the detection mechanism. In the next stapes of this research the detection mechanism will be improved to detect more sophisticated kinds of Jammers.

### Acknowledgment

This research has been partially funded by the Federal Ministry of Education and Research of Germany under grant numbers 16ES1131 and 16ES1128K.